\documentclass[conference]{IEEEtran}
\ifCLASSINFOpdf
\else
\fi
\usepackage{hyperref}
\usepackage{cite}
\usepackage[framemethod=tikz]{mdframed}

\usepackage{supertabular}
\usepackage[textsize=tiny]{todonotes}
\usepackage{graphicx}
\usepackage[threshold=100,thresholdtype=words,autopunct=false]{csquotes}

\begin{document}
\title{Why People Still Fall for Phishing Emails: An
Empirical Investigation into How Users Make Email
Response Decisions}


% \author{
% \IEEEauthorblockN{Asangi Jayatilaka}
% \IEEEauthorblockA{Centre for Research on Engineering \\Software Technologies (CREST), \\The University of Adelaide, Australia\\
% asangi.jayatilaka@adelaide.edu.au}
% \IEEEauthorblockA{School of Computing Technologies \\ RMIT University, Australia\\
% XXXXXXXXXXX}
% \and
% \IEEEauthorblockN{Nalin Asanka Gamagedara
% Arachchilage}
% \IEEEauthorblockA{School of Computer Science,\\ The University of Auckland, New Zealand \\
%  nalin.arachchilage@auckland.ac.nz}
% \and
% \IEEEauthorblockN{M. Ali Babar}
% \IEEEauthorblockA{Centre for Research on Engineering \\Software Technologies (CREST), \\The University of Adelaide, Australia\\
% ali.babar@adelaide.edu.au}
% }

\author{
    \IEEEauthorblockN{
                    Asangi Jayatilaka\IEEEauthorrefmark{1}\IEEEauthorrefmark{2}, 
                    Nalin Asanka Gamagedara Arachchilage\IEEEauthorrefmark{3},
                    Muhammad Ali Babar\IEEEauthorrefmark{1}
    }
    \IEEEauthorblockA{
    \IEEEauthorrefmark{1}Centre for Research on Engineering Software Technologies (CREST),
     The University of Adelaide, Australia\\
     \{asangi.jayatilaka, ali.babar\}@adelaide.edu.au \\
    \IEEEauthorrefmark{2}School of Computing Technologies, RMIT University, Australia\\
     asangi.jayatilaka@rmit.edu.au\\
    \IEEEauthorrefmark{3}School of Computer Science, The University of Auckland, New Zealand \\
    nalin.arachchilage@auckland.ac.nz
    }
}%end author


%Telephone: (800) 555--1212, Fax: (888) 555--1212}


\IEEEoverridecommandlockouts
\makeatletter\def\@IEEEpubidpullup{6.5\baselineskip}\makeatother
\IEEEpubid{\parbox{\columnwidth}{
    Symposium on Usable Security and Privacy (USEC) 2024 \\
    26 February 2024, San Diego, CA, USA \\
    ISBN 979-8-9894372-5-2 \\
    https://dx.doi.org/10.14722/usec.2024.23xxx \\
    www.ndss-symposium.org, https://www.usablesecurity.net/USEC/
}
\hspace{\columnsep}\makebox[\columnwidth]{}}

\maketitle

\begin{abstract}
Despite technical and non-technical countermeasures, humans continue to be tricked by phishing emails. How users make email
response decisions is a missing piece in the puzzle to identifying why people still fall for phishing emails. We conducted an empirical
study using a think-aloud method to investigate how people make ‘response decisions’ while reading emails. The grounded theory
analysis of the in-depth qualitative data has enabled us to identify different elements of email users’ decision-making that influence
their email response decisions. Furthermore, we developed a theoretical model that explains how people could be driven to respond to
emails based on the identified elements of users’ email decision-making processes and the relationships uncovered from the data. The
findings provide deeper insights into phishing email susceptibility due to people’s email response decision-making behavior. We also
discuss the implications of our findings for designers and researchers working in anti-phishing training, education, and awareness
interventions.
\end{abstract}


% \fi
%%\IEEEpeerreviewmaketitle

\section{Introduction} \label{introduction}
%-------------------------------------------------------------------------------

Phishing is one of the most prominent and influential cyber attacks, as it is a precursor for many other attacks, including identity theft, which is among the worst \cite{Rub4, trends}.  Phishing attacks have sharply risen recently, partly driven by COVID-19 and supply chain uncertainty \cite{abroshan2021covid}. %%
Phishing activity trends report of the Anti-Phishing Working Group reports more than 4.7 million attacks for 2022.
This report also highlights that since the beginning of 2019, the number of phishing attacks has grown by more than 150\% per year.  Phishing attacks are most commonly launched through emails as they are difficult to detect~\cite{stats, salloum2022systematic}. Often phishing email attacks target banks, defense organizations, and private companies, as they curate a wide variety of data, including personal and financial data~\cite{bose2014phishing}. Generally,  clicking on links, downloading attachments, or replying to phishing emails can be considered unsafe response decisions \cite{lawson2020email, parsons2013phishing}.
CISCO’s 2021 Cybersecurity threat trends report states that at least one person has clicked a phishing link in around 86\% of organizations \cite{CISCO}.  A successful phishing attack can trick users into unintentionally disclosing their valuable information, compromising their devices or accounts \cite{smith2019computer}. Additionally, phishing attacks are also used for installing malware (i.e., malicious software), which can disturb the normal operations of a computer system, contributing to significant financial losses and reputation damages.

Technology alone is insufficient to combat phishing email attacks; therefore, transforming users from the weakest line of defense to the most robust line of defense is essential~\cite{franz2021sok, gupta2018defending, pilavakis2023didn}.  
An active community of practitioners and researchers focuses on phishing education, training, and awareness (e.g., phishing alerts) to support users in correctly identifying phishing~\cite{desolda2021human}. However, these efforts are with limited success~\cite{allodi2019need, franz2021sok}. A major challenge in designing effective anti-phishing interventions is the lack of attention to reasons why people still fall for phishing  \cite{franz2021sok, wash2020experts, kirlappos2011security, moreno2017fishing}.

Understanding why people still fall for phishing emails will provide underpinning science to design effective future anti-phishing tools and educational interventions. Although prior literature largely focuses on analyzing the personality and demographics of people who fall for phishing emails \cite{sheng2010falls, lin2019susceptibility}, how people make email responses and the thought process a user goes through when deciding
how to respond to their emails is often overlooked~\cite{jayatilaka2021falling, franz2021sok, wash2020experts, kirlappos2011security, moreno2017fishing}.
Only a limited number of studies have attempted to conduct qualitative user studies to explain people’s email decision-making processes \cite{wash2020experts, wash2018provides, jayatilaka2021falling, wash2021knowledge, pilavakis2023didn}. Although qualitative studies, compared to quantitative studies,  allow us to obtain more detailed and holistic insights into users' email response decision-making behaviors and the reasons for those behaviors, even such prior work does not interpret the different elements of people’s email response decision-making processes and their relationships influencing their email response behavior.

To address this research gap, we conducted an empirical investigation through a ``think-aloud" role-play experiment and follow-up interviews to better understand people's decision-making behavior when responding to emails. We developed a theoretical model that explains how people are driven to respond to emails by clicking on links, replying, and downloading attachments based on the identified elements of the people's email response decision-making process and the relationships uncovered from collected data. The model developed based on empirical evidence interprets how different elements of people’s email response decision-making processes could positively and negatively influence people's intention to respond to emails, which was lacking in previous literature. For example, the model provides deeper insights into how certain habits, validation techniques, previous experiences, etc., can positively influence people's intention to respond to emails, as a result increasing the risk of them falling for potential phishing attacks.
In summary, our contributions are as follows:

\begin{itemize}

 \item Based on empirical evidence and grounded theory analysis, we provide knowledge into elements of email users' decision-making process (e.g., diverse types of emotions and personal habits) that influence their email response decisions.

\item We develop a theoretical model that explains how people are driven to respond to emails by clicking on links, replying, and downloading attachments based on the identified elements of the user's email response decision-making process and their relationships uncovered from data.
The developed theoretical model provides deep insights (i.e., scientific underpinnings) into an individual’s email decision-making process or response behaviors to emails in general. As a result, the model enables us to identify general email decision-making flaws that attackers could potentially exploit to launch successful phishing attacks. Furthermore, understanding people's general email decision-making flaws the attacker could leverage may help better design technical countermeasures such as anti-phishing tools to thwart email-based phishing attacks.   
\end{itemize}

\section{Background and motivation} \label{relatedwork}

%Despite the advances in anti-phishing education, training, and awareness interventions \cite{franz2021sok, misra2017phish, sheng2007anti, gordon2019evaluation, gupta2018defending, wen2019hack, dixon2019engaging},  people still increasingly fall prey to phishing emails \cite{gupta2018defending}. 

%%To better understand how any human-centric anti-phishing intervention can be designed, an in-depth understanding of users' susceptibility to phishing attacks is crucial~\cite{CHI2020}. Unfortunately, existing human-centric anti-phishing interventions often lack a scientific basis that provides a detailed knowledge of psychological and behavioral aspects related to why phishing emails still work \cite{wash2020experts, franz2021sok, kirlappos2011security, moreno2017fishing, ge2021personal, wang2017coping}.

It is imperative to obtain a deeper understanding of the scientific underpinnings of how users make email response decisions to design better technical and non-technical countermeasures that thwart email-based phishing attacks.
Research on phishing email susceptibility has considered the demographic or personality of victims \cite{lawson2020email, jones2019email,  ge2021personal}, %aldawood2018educating, cho2016effect
or phishing email characteristics \cite{lin2019susceptibility, caspi2022effects, ferreira2019persuasion, williams2019persuasive}.  
Conversely, several studies have investigated psychological and behavioral responses in this regard \cite{williams2019persuasive, greene2018user, parsons2013phishing}; they mostly employ existing theories borrowed from other fields to derive phishing models \cite{moody2017phish, shahbaznezhad2021employees, wang2016overconfidence, kwak2020users, goel2017got, molinaro2018evaluating, vishwanath2011people, wang2017coping}. Such work often employs hypothesis-testing to see if the theory, specified as the hypothesis, is supported by gathered quantitative data. 

While models derived from existing theories and validated with quantitative data provide much-needed insights into phishing, they often focus only on specifically selected dimensions of email response behaviors depending on the selected theories. For example, researchers in \cite{molinaro2018evaluating} focused on the phishing cues available in the emails. %%
They hypothesized that phishing cues are linearly combinable and hence a type of ``Judgment Analysis", is appropriate for evaluating phishing judgments. 
Through role-play, experiments conducted with participants who judged whether emails were phishing proved that their hypothesis was correct.
In another study~\cite{goel2017got}, researchers focused on the impact of the content and framing of phishing emails on user vulnerability.   They came up with several hypotheses and evaluated those using surveys with University students. Their results suggest that the desire to protect things of value and the opportunity to obtain valued objects could make people vulnerable to phishing attacks.
Researchers in \cite{shahbaznezhad2021employees} derived several hypotheses based on the protection motivation theory and the theory of planned behavior, focusing on the individual, organizational, and technological factors that affect phishing email responses.  They later tested the validity of the derived hypothesis using questionnaires. Inspired by research on information process and interpersonal deception, researchers in \cite{vishwanath2011people} developed an integrated information processing model of phishing vulnerability. % Researchers in \cite{vishwanath2011people} developed an integrated information processing model of phishing vulnerability
% % , with 22 hypotheses,  % based on prior research on information process and interpersonal deception. 
Upon validating the model with undergraduates, the researchers found attention to the email source, grammar and spelling, urgency cues, and subject line were significantly negatively related to an individual's likelihood of responding to a phishing email.

Only a limited number of studies have conducted qualitative user studies to propose theoretical models to explain people's email decision-making processes  \cite{wash2020experts, wash2018provides, jayatilaka2021falling, wash2021knowledge}.   For example,  Wash \cite{wash2020experts} conducted interviews with experts to identify the process that they follow to identify phishing messages they received in the past.  They found that experts follow a three-stage process for identifying phishing emails  \textemdash  \, first sense-making, then
suspicion, then acting. On the other hand,  Jayatilaka et al. \cite{jayatilaka2021falling} revealed eleven high-level themes influencing people's email response behaviors \cite{jayatilaka2021falling}. However, the study \cite{jayatilaka2021falling}  failed to interpret the elements that constitute these themes and the nature of their relationships leading to the email responses.

In addition to the above-mentioned gaps, several other common drawbacks that exist in most aforementioned studies are described below. Firstly, it is common for images of emails to be used in phishing-related experiments \cite{reinheimer2020investigation, cui2020effects, zheng2022presenting}; however, this could detach participants from their naturalistic setting, affecting their decision-making. 
Moreover, such images prevent understanding whether or not people use the link URL information shown in the status bar without alerting participants~\cite{ parsons2013phishing, wang2016overconfidence}. 
Secondly, surveys are often utilized to collect data from the users about their habits, traits, and explanations for decisions they have made~\cite{nasser2020role, harrison2016individual, ackerley2022errors, vishwanath2011people, wang2017coping, chen2020examination, wash2021knowledge, wang2017coping}.
However, such self-reported surveys may not provide adequate information or be inaccurate as users could justify the corresponding email response. Jaeger et al. \cite{jaeger2021eyes} have used eye-tracking software in their experiment to reduce the limitations of using a survey. 
Thirdly, most research request participants to make pre-defined email legitimacy judgments or response decisions  (e.g., phishing or legitimate)~\cite{sarno2021so, lawson2017interaction, jones2019email, wang2016overconfidence, wen2019hack, reinheimer2020investigation, zheng2022presenting, butavicius2022people}; however, this may not align with how people behave naturally. 
For example, our study findings reveal that, at times, people could develop doubts about email legitimacy and hence make a final decision only after validating the email. Fourthly, several research studies have relied on participants' descriptions of past incidents/situations to understand how they detect phishing emails \cite{wash2020experts, wash2021knowledge}. Such methods could suffer from the imperfect memory of participants of past incidents. 

The current study aims to provide deeper insights into users’ email response decision-making processes by developing a theoretical model explaining how people are driven to respond to emails by clicking on links, replying, and downloading attachments based on elements of the user's email response decision-making process and their positive and negative relationships uncovered from grounded theory analysis of qualitative data.  To gather qualitative data about participants’ email response decision-making behaviors we employed a role-play-based think-aloud method and follow-up interviews using a simulated email client. The reasons for this data collection method and set-up are further explained in Section~\ref{methods}. Armed with this deeper understanding of email response decision-making elements and their relationships, we can begin to design more effective anti-phishing tools and techniques to face phishing attacks successfully.

\section{Research Methodology} \label{methods}

\begin{figure}[!htb]
    \centering
    
        \includegraphics[width=0.85\linewidth]{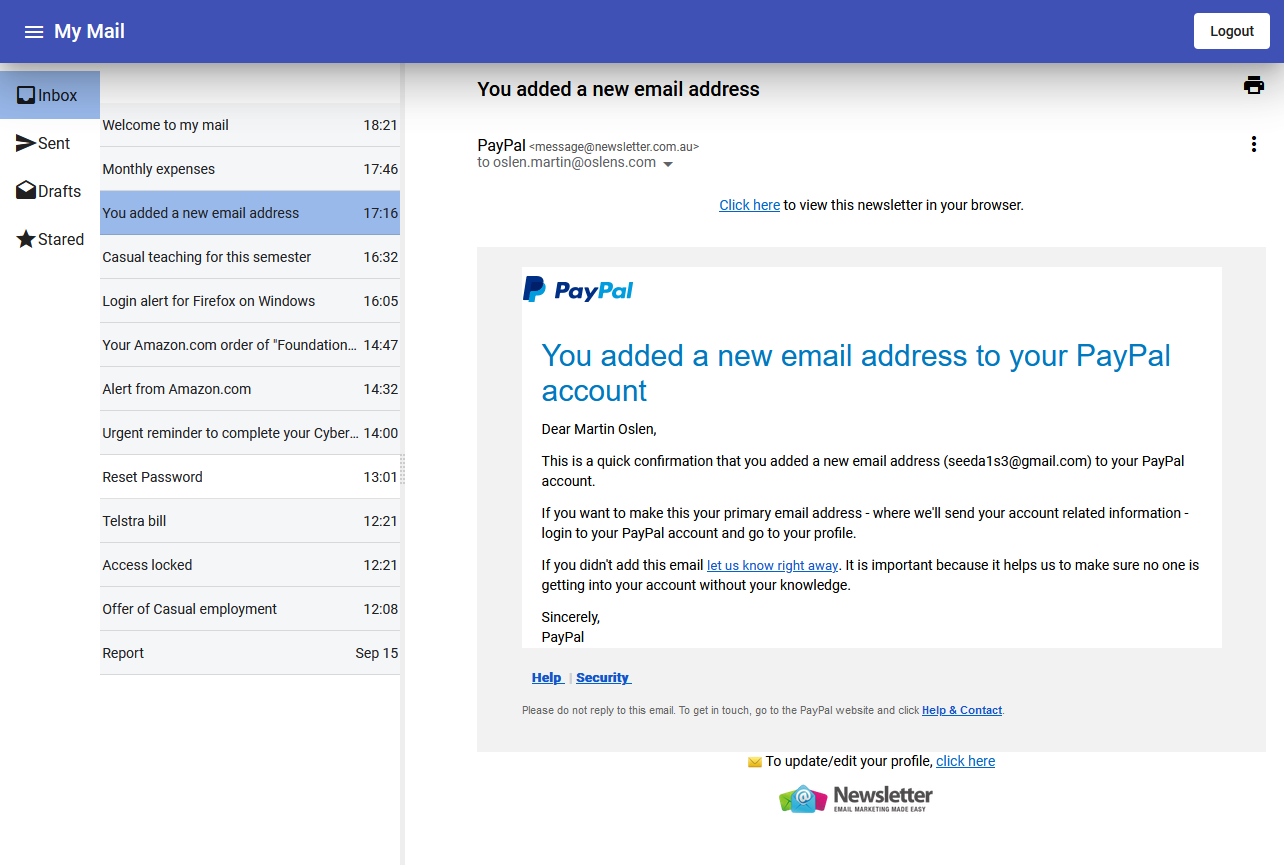}
        \caption{Simulated  web  email  client}
        \label{emailclient}
\end{figure}

We decided to use a  role-play-based think-aloud method to achieve the aim of this research instead of surveys ~\cite{nasser2020role, harrison2016individual, ackerley2022errors, vishwanath2011people, wang2017coping, chen2020examination, wash2021knowledge} or retrospective interviewing \cite{wash2020experts, wash2021knowledge}, as those: i)~rely on memory of past incidents, ii)~don't provide detailed insights into the decision-making processes, or iii)~rely on people's justifications for email responses already determined. In our study, participants were not requested/forced to make certain decisions (e.g., classifying emails as legitimate/phishing), allowing them to explain their intention to respond to emails naturally. Having a simulated email client (see Figure~\ref{emailclient}) enables the participants to naturally engage with the activity while contextualizing their opinions and explaining how they would respond to emails by recalling their memories in a naturalistic environment.  We conducted follow-up interviews to expand on the think-aloud results. As the emails were embedded in the simulated email client, we were able to automatically adjust the dates and times specified in the email content to suit the time and day that a participant carried out the experiment.  The research was approved by the ethics committee of the University of Adelaide.   

 In this study, in terms of email responses, we focus on clicking on links, replying, and downloading attachments because, as mentioned before, they are considered unsafe email responses in the context of phishing. Henceforth, for brevity, we refer to clicking on links, replying, and downloading attachments as email responses throughout this paper.

%\begin{figure}
%\centering
%\includegraphics[trim={10px 250px 150px 0px },clip,width=0.6\linewidth]{images/TheototicalSaturation2.png}
%\caption{The number of unique concepts for each participant}
%\label{saturation}
%\end{figure}

 \begin{figure}[!htb]  
   
        \centering
        \includegraphics[trim={10px 250px 150px 0px },clip,width=0.78\linewidth]{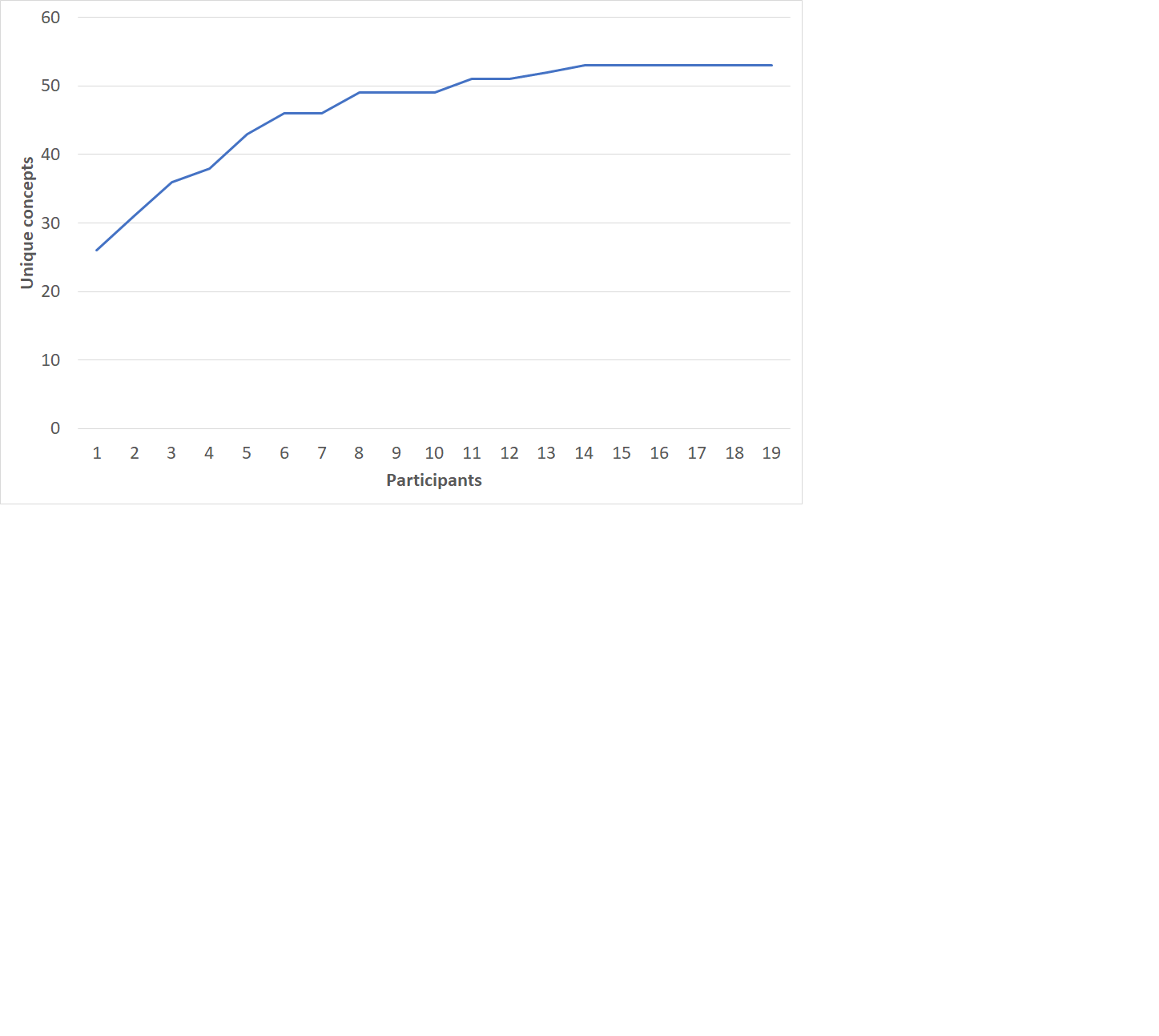}
        \caption{The number of unique concepts for each participant}
        \label{saturation}
  
\end{figure}

\subsection{Emails  selection }

We used real phishing emails (12) and corresponding real legitimate emails (12) for this study\footnote{Screenshots of the phishing and legitimate emails used for the study are included in this link: \url{https://osf.io/gu7qs/?view_only=1aa1e0ab5e47440aa5cb4e7a6a49d9c8}}.   As we aim to obtain a holistic picture of how people are driven to respond to emails, the selected emails reflect different domains (e.g., education, financial, social media), and attacker strategies. Examples of emails include  University password expiry,  Facebook alerts,  Amazon order confirmations, and mobile bills. Several phishing strategies, such as providing a personal salutation,  creating fear and urgency, mimicking the appearance of a legitimate email, mimicking the sender’s email address, including the signature of the sender, and URL obfuscation, highlighted in the literature that is used to to create fraudulent emails were included in the selected list of emails \cite{misra2017phish, stojnic2021phishing}.  The phishing emails were sourced from various venues \cite{USberkerly, sensorstech, scamdetect}. The emails were adapted to the scenario (e.g., the name of the recipient was changed to the name of the fictitious character in the role-play). The corresponding legitimate emails were sourced from researchers' email correspondences (e.g., if the phishing email is about a shared Google spreadsheet,  the corresponding legitimate email was also chosen to be about a shared Google spreadsheet).  Based on the email assignment process which is described at the end of this sub-section,  a  participant did not receive a legitimate email and its corresponding phishing email together in his or her inbox. 

Real phishing links from PhishTank~\cite{PhishTank} were adapted as phishing URLs for this study. Recent research points to six  URL obfuscation techniques  \cite{garera2007framework, fernando2020johnny}. We included two phishing emails for all except one URL obfuscation technique (i.e., obfuscating with HTTPS schema) pointed out \cite{fernando2020johnny}. For obfuscating with HTTPS Schema to occur, the browser must navigate to the destination (i.e., a landing page) to show the green padlock icon on the address bar. However, as previous studies have indicated, most people clicking on phishing links will go on to disclose information to those phishing websites \cite{sheng2010falls}. Therefore, as our work focuses on understanding how people make email response decisions while reading their emails, the browser-based email client was prevented from navigating to any URL destination when a participant clicked on an email link. Instead, regardless of its legitimacy, a “link clicked”  message was shown to the users when they clicked on any email link. Therefore, the obfuscation technique with HTTPS Schema is not meaningful in our study \cite{fernando2020johnny}. Most research on phishing emails focuses on emails with links neglecting emails without links; however, as explained before, phishing emails can come as emails without links. Therefore, we included several emails without any links. This included two emails requesting download attachments and two emails requesting to reply. 

When assigning the emails, we selected 12 emails from the pool of emails.  Then, considering a phishing email and the corresponding legitimate email as a pair,  we randomly assigned the legitimate email to 50\% of the participants and the phishing email to the remaining 50\%.  Following this approach we ensured that a given participant will not receive a legitimate email and its corresponding phishing email. Also, given an email pair, 50\% of the participants will receive legitimate, and the remaining 50\%  will receive the corresponding phishing email to reduce any biasness in the analysis.

 \subsection{Data collection}
We recruited 19 students from the University of  Adelaide by distributing flyers.  Eight participants were male and 11 were female. Twelve were undergraduate students and 7 were post-graduate students. They were distributed across the Faculty of Arts (5), Faculty of Health and Medical Sciences (5), Faculty of the Professions (4), and Faculty of STEM (5). Fourteen participants were 25 years of age and younger,  and  5 were over 25.  Six participants mentioned that they have had some form of anti-phishing training. Out of these six participants, 4 were undergraduates and 2 were post-graduates. Each data collection session was conducted over Zoom and lasted for around 90 minutes. The participants were requested to share their screens with the researcher so that the researcher was able to better understand how the participants interacted with the emails. These Zoom sessions were recorded and later used to generate the transcripts for the analysis.

Participants were asked to assume they were a fictitious person named ``Martin Oslen”.   The background information on Martin was provided to the participants through a document.  The fictitious person (``Martin") attends the participants' University, and the scenario included information on Martin's: i) social and professional interactions with their contacts (e.g.,   supervisor details, the mobile service provider, wife's details); and ii) online services and platforms (e.g., details about his social media,   platforms he uses for online shopping such as Amazon and his banking details). We asked participants to role-play and debrief Martin’s scenario before opening his email inbox.  We allowed them to retain and refer to the document whenever they needed to throughout the study.  When the participants were comfortable with the scenario they were given the URL of the email client and were asked to share their screens. We then provided basic descriptions of the email client to the participants. Furthermore, before the think-aloud session,  participants were given a practice email to get familiar with the setup.

During the think-aloud session, we requested the participants explain how they felt to receive each of the given emails and how they would react to each one of them.   We asked participants to talk freely about anything that came to their minds when going through the emails. We explained to them that the goal of the study is to obtain deeper insights into how they make email responses usually and not to evaluate the appropriateness of those decisions or responses. Therefore, they were requested to act as they would normally do for the given emails.
In this process, we made sure not to request participants to perform certain actions (e.g., instructing participants to hover their mouse over a link to see the destination URL  in the status bar~\cite{hakim2021phishing}) or make certain decisions (e.g., classifying whether an email is legitimate or phishing) as that could influence their usual behavior.  The follow-up questions were asked based on what participants explained in the think-aloud session to further clarify their thought processes.

Two pilots were conducted which allowed the researchers to identify and introduce three changes to how the think-aloud sessions and the follow-up interviews were conducted.  
During the first pilot,  we identified that the initial question we asked during the think-aloud session (i.e., ``how legitimate do they think a given email is”) restricted the participants from showing their natural behaviors. That question forced them to first make a decision on the email’s legitimacy without much consideration and later think of justifications to support their decisions. We decided to replace this question during the second pilot session with ``how do you feel about receiving this email”. This allowed the participants to behave more naturally and openly discuss what came to their minds and later explain the desired actions. Secondly, the pilot interviews also demonstrated the importance of providing a practice email to participants so they can be familiarized 
with the setup. Thirdly, we asked follow-up questions after each email in the second pilot as participants faced difficulties answering the questions when follow-up questions were asked at the end of all the emails in the first pilot. We observed that the approach used in the second pilot was more convenient for the participants to answer the questions while their memory was still fresh.

\subsection{Data analysis}\label{dataanalysis}

We used Grounded Theory (GT) \cite{strauss1998basics, glaser1967discovery} for data analysis, using NVivo\textsuperscript{TM} software.  % The first author conducted the data analysis, and the second author verified all the codes throughout the process to reduce bias and increase the reliability of the findings. Similar to previous qualitative studies \textcolor{red}{ref}, the categories and their relationships were thoroughly discussed and finalized after several revisions among all authors. 
The data analysis included different types of coding: \textbf{\textit{open coding}}, \textbf{\textit{axial coding}} and \textbf{\textit{selective coding}} \cite{strauss1998basics}. During \textbf{\textit{open coding}}, we read through the interview transcripts line by line and encapsulated them into codes with short phrases. Later we performed a second round of open coding to refine the results of the first round. We coded the interviews based on contextualized statements instead of single terms. After coding, we discussed the relations between the newly found codes and agreed upon a set of higher-level axial codes  (i.e., \textbf{\textit{axial coding}} and these will be called concepts hereafter). The concepts served as a baseline for the next round of coding, which is \textbf{\textit{selective coding}}. During the selective coding process, all researchers agreed upon a set of codes (called core categories hereafter) that represent the different elements that affect the employees’ email response behaviors. The first author led the data analysis, and the second author reviewed the emerging codes, concepts, and categories along with the interview transcripts during each step of the data analysis process; any disagreements were resolved through discussion before moving further into the analysis. Frequent iterative group meetings with all the authors were held throughout the process to ensure data interpretation consistency and rigor. All three authors have extensive experience with various qualitative methods including grounded theory and thematic analysis.  We achieved saturation through this process (i.e., where no fresh information emerges from subsequent think-aloud sessions) because all the main categories and concepts had been uncovered across the 19 participants (see Figure~\ref{saturation}).

The last step of our GT approach was to form a theoretical model by considering the relationships between the discovered axial and selective codes.  Similar to a previous study \cite{mai2020user}, the model was based on the knowledge (i.e., categories and concepts) that emerged from grounded theory analysis.   Model relationships were identified through rigorous and iterative analysis of transcripts based on “interpretive data” of people’s email response behaviors. 
All researchers discussed the draft theoretical model to reach an agreement.  We ensured the completeness and accuracy of the model in terms of the collected data through negative case analysis~\cite{mai2020user, brodsky2008negative}. Here we went through interviews iteratively to check whether the participants’ statements could be assigned to the draft theoretical model. If not, we identified how they diverged from our draft and adopted it accordingly. We iteratively refined our theoretical model until all statements were captured.

It is important to note that, as the goal of the study was to explore individuals' decision-making processes that influence their email response decisions,  all concepts and relationships that emerged from the qualitative data (involving 19 participants where each participant went through 12 emails during the study) are included in the model regardless of their frequencies.  Therefore,  not all categories and concepts are reflected in all our participants' email response decision-making processes. As quantitative results in qualitative research cannot be used to generalize findings; hence we will discuss all statements in Section~\ref{sec:results} without providing numbers.

%The developed theoretical provides deeper insights into individual's email decision-making process or response decisions to emails in general. As a result, the model enables us to identify general email decision-making flaws that attackers could potentially exploit to launch successful phishing attacks. 
%rather than only focusing on flaws related to phishing emails.}

\section{Results} \label{sec:results}

\begin{figure*}
\centering
\includegraphics[trim={0px 0px 0px 0px },clip,width=0.57\linewidth]%{images/OverviewModel.drawio (5).drawio.png}
{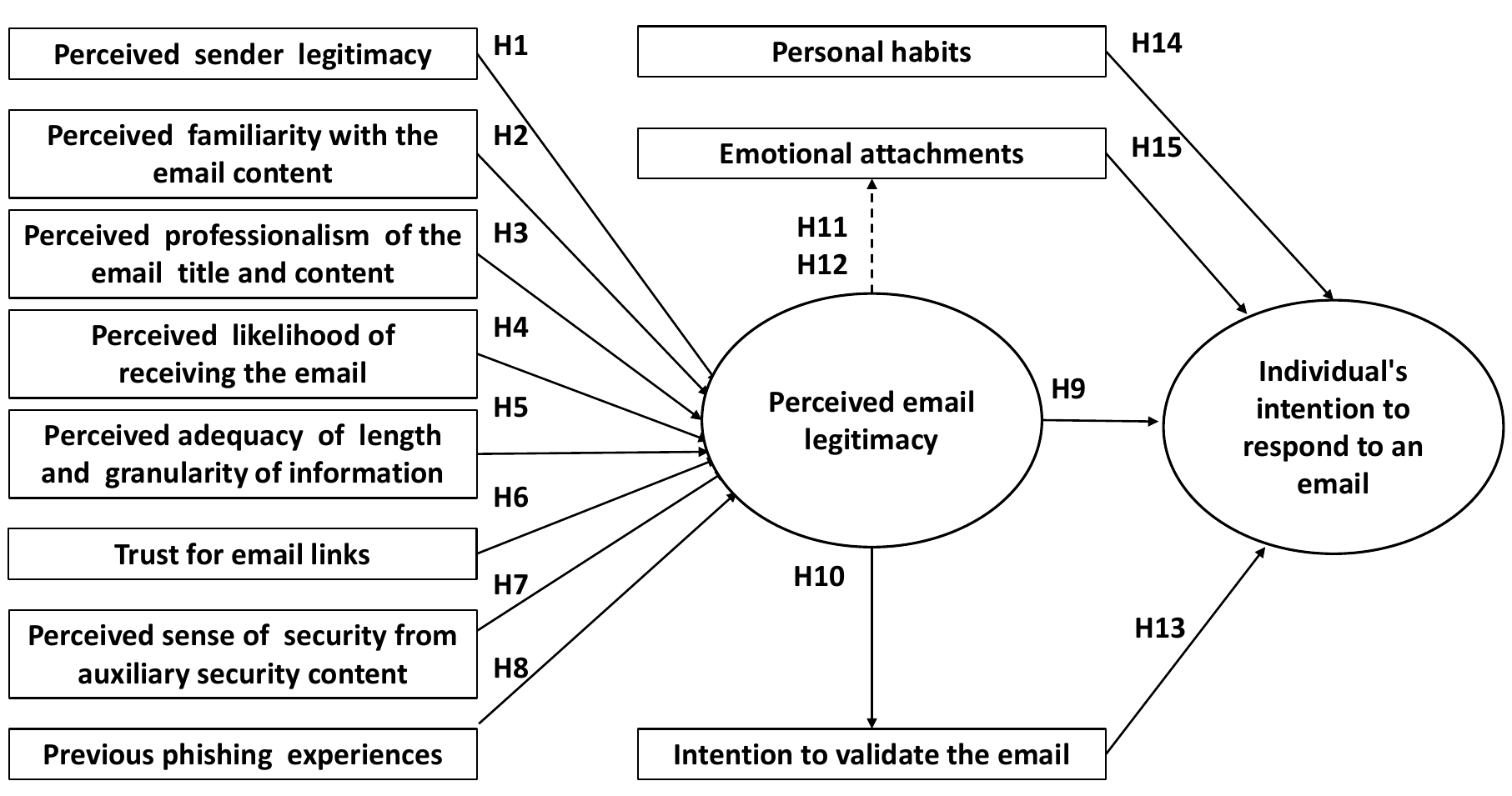}
\caption{ Overview of the theoretical model developed based on gathered qualitative data. The
latent variables are indicated with ovals, and core categories
derived based on GT analysis are shown in rectangles. Concepts are not shown in the figure for brevity. In a  `partially affect' relationship only some concepts in a category are affected by the other category or latent variable. }
\label{modeloverview}
\end{figure*}

%Grounded theory analysis of the collected data revealed 51 concepts attributed to 11 core categories that influence email response behaviors. 
An overview of the theoretical model that emerged from the data is presented in Figure~\ref{modeloverview}. 
In summary,  an individual's email response behavior is determined by how they perceived the email legitimacy, which in turn is influenced by the concepts of perceived sender legitimacy, perceived familiarity, perceived professionalism in the email title and body, perceived likelihood of receiving the email, perceived adequacy of length and granularity of information, trust for email links, perceived sense of security from auxiliary security content and previous phishing experiences. Furthermore, an individual's intention to respond to an email is also determined by their emotional attachments and personal habits, as well as their intention to validate the email.  In-depth knowledge about diverse concepts that constitute the high-level categories and the nature of their relationships leading to email responses is provided below.  %For example,  different personal habits, emotions,  and email validation techniques could positively and negatively affect people's intention to respond to an email. 

\subsection{H1: Effects of ``perceived  sender  legitimacy"  on  ``perceived email legitimacy"} \label{senderlegitimacy}

 Our findings reveal that how people perceive the sender's legitimacy is pivotal in the trust they place in a received email. People tend to focus on specific components of the sender's email address, sometimes, without even having a clear understanding of which components need more attention. Given the sender's address, most participants fail to place more emphasis on the domain than the sub-domain, email display name, and sender's user name. One participant explained that he is convinced that a given email is coming from a reliable source by only considering the sender's email display name overlooking the issues he observed in the sender's domain \blockquote[P15--P]{It is a bit dodgy. But I'm convinced it's from PayPal [pointing to the sender's email name].} [\textcolor{blue}{see H1.1}].  Some participants made decisions about the sender's legitimacy by looking at the sender's user name. Having no-reply as the user name, having a familiar or professional user name allows a user to have more faith in the sender [\textcolor{blue}{see H1.2}]. Also, participants focused on the domain of the sender address in deciding the sender's legitimacy [\textcolor{blue}{see H1.3}]. We often notice when the sender address domain is deemed to be known or looked professional, participants tend to believe the sender is legitimate. In some instances, we observed that participants get confused with the sub-domain and domain. They, at times, cannot differentiate the domain and sub-domain. A participant explained that she is not trusting the sender as she was not confident about the sub-domain of the sender's email address even when she was confident about the domain  \blockquote[P02--L]{I don't think that they have edm [pointing to the sub-domain] in the bank emails. It doesn't make sense} [\textcolor{blue}{see H1.4}].

A few participants looked into the reply-to-address to draw conclusions about the sender's legitimacy. For example, one participant got suspicious after seeing the reply-to address and identifying that the reply could be going to a Gmail account instead of the  sender's official address \blockquote[P06--P]{When I reply, it will go to this Gmail, I would delete my email}. On the other hand, there was another participant who suspected the sender's address; however, after looking at the reply-to address (the user name of the reply-to address), she was much satisfied with the sender's legitimacy  \blockquote[P01--P]{This email is there [mary@sau23.org in as the reply-to email]. Mary at dot org} [\textcolor{blue}{see H1.5}]. We also observed situations where participants had issues trusting the received emails after noticing the spelling issues in the sender's email address 
%\blockquote[P16--P]{If I notice that anything [in the sender's email address]  off by one letter or something like that, that alone can just make me not want to read the email what so ever}
[\textcolor{blue}{see H1.6}].

People also look at the sender's full email address  to make conclusions about  the sender's legitimacy. If the sender's address is known or familiar to the participants, then they seem to instantly develop trust about the sender \blockquote[P16--P]{That's like a staff email address from my experience. I feel good} [\textcolor{blue}{see H1.7}]. Some participants made assumptions about the sender based on the email addresses specified in the email body [\textcolor{blue}{see H1.8}]. For example, a  participant was satisfied with the sender's legitimacy after seeing the address specified in the body of a shared google sheet  and overlooked the issues that he observed in the sender's email address specified in the email header  \blockquote[P09--P]{It looks like just a shared thing from the wife [looking at the email adders in the email body]}. Therefore, based on the data, the following relationships exist: \\

%Mention of theories around misspelling and grammar errors in phishing emails in connection to H1.6

% \vspace{1em}
\begin{mdframed}
\textbf{H1.1 - H1.5:} Perceived legitimacy of the [sender's email display name, sender address user name, sender address domain,  sender address sub-domain, reply-to address] positively affects perceived email legitimacy\\
\textbf{H1.6:} Spelling issues in the sender email  negatively affect perceived email legitimacy\\
\textbf{H1.7 - H1.8:} [Knownness of the sender address in the email header, knownness of the sender addresses specified in the email body]   positively affects perceived email legitimacy
\end{mdframed}

\subsection{H2: Effects of ``perceived  familiarity of the email content"   on  ``perceived email legitimacy" }
We observed people often intentionally and unintentionally compare the received emails with what they have seen before.  More specifically, our results  explain which aspect of emails people particularly consider when making judgments about the familiarity of the email content. Based on the data  people consider the familiarity of the writing style [\textcolor{blue}{see H2.1}], the familiarity of the interface elements [\textcolor{blue}{see H2.2}], and the familiarity with email layouts [\textcolor{blue}{see H2.3}] when making judgments about email legitimacy. For example, one participant explained her reasons for trusting a given phishing email \blockquote[P05--P]{It's similar to the emails that  I know to be real emails from the University Human Resources. So,  I already trusted the layout in my head}. There were situations where participants made decisions about email legitimacy only based on their perceived familiarity with writing styles, interface icons, or email layouts. For example, \blockquote[P18--L]{Okay, it's a Google Sheets looks very fine. I don't think I wouldn't check the sender's address in this one}. Hence based on our data, the following relationships exist: \\

% \vspace{1em}
\begin{mdframed}

\textbf{H2.1 - H2.3:} Familiarity with the [writing style, interface elements, email layout]  positively affects perceived email legitimacy
\end{mdframed}

\subsection{ H3: Effects of ``perceived  professionalism  of the email  title and content"   on ``perceived email legitimacy"}
 Our data reveals insights into  specific aspects  (e.g.,  email layout [\textcolor{blue}{see H3.1}], writing style  [\textcolor{blue}{see H3.2}], and interface elements [\textcolor{blue}{see H3.3}]) people tend to focus in terms of professionalism in emails. 
With respect to the writing styles,  participants looked into the professionalism in the writing styles of the email title and email body. For example, a participant explained that she is not trusting a given legitimate email because she is not convinced that the title of the email is professional enough \blockquote[P01--P]{Because it should have a kind of professional way like Notification of Declining Transaction something like that}.
In terms of layout, participants considered how the email is organized, including the layout of the email links. One participant explained seeing other links such as `contact us in an email makes them trust those emails \blockquote[P01--P]{They have these contact links, which would certainly be in a reliable source. So, I think including these links is really frequent in, you know, in a trusted company}. 
In terms of interface elements,  participants explained that they look for professionalism in color schemes, buttons, font styles, and logos used in the email. There were instances where participants got suspicious about email legitimacy, even for a legitimate email, when they thought the email didn't have professional-looking interface elements, for example, buttons \blockquote[P14--L]{I  don’t look at anything else. After seeing that [button for the link], I am pretty much sure that it's a phishing email}. Therefore, based on our data, the following relationships exist: \\

% \vspace{1em}
\begin{mdframed}
\textbf{H3.1 - H3.3:} Perceived professionalism in the [email layout, writing style, interface elements] positively affects perceived email legitimacy
\end{mdframed}

\subsection{H4: Effects of ``perceived  likelihood of   receiving the email"   on ``perceived email legitimacy"}
%Previous research has highlighted that alignment of the email context with user context is a significant factor in how people would respond to emails~\cite{greene2018user, goel2017got}. Our data confirm this. 
People often trust an email in situations where they are expecting such an email [\textcolor{blue}{see H4.1}]. For example, a participant mentioned that they could be quite confident about an  Amazon order confirmation email if he knew that he had placed such an order \blockquote[P02--P]{I think I would be very confident getting this email if I know that I have placed an order. I think this is kind of an expected email}. We also observed that participants even tend to ignore phishing cues they observe in emails if they perceive the likelihood of receiving those emails is high. \\

% \vspace{1em}
\begin{mdframed}
 \textbf{H4.1:}  Perceived expectancy of the email positively affects perceived email legitimacy
\end{mdframed}

\subsection{H5: Effects of ``perceived adequacy  of  length and  granularity of information"   on ``perceived email legitimacy"}

Our findings reveal that, at times, people tend to look into the length and granularity of the email when making email legitimacy judgments. There seems to be a  misconception that phishing attackers would not create lengthy emails or will not have the capacity to access their granular personal information (e.g., name, account numbers, etc). 
In terms of the length of an email [\textcolor{blue}{see H5.1}], P1 explained that she does not think a phishing attacker would create long emails to make people fall for those emails, so she has much trust in lengthy emails \blockquote[P01--P]{This content being a really long email, I think they are actually trying to convey me this message, not a spam-like ... I think a spammer doesn't want to type this much to make a person fall into trap}. 
On the other hand, when the email has detailed information about the situation and/or information specific to the  participants, they tend to believe it is a legitimate email [\textcolor{blue}{see H5.2}]. For example, a participant explained,  \blockquote[P01--L]{I think this is a legitimate email because it includes an item that I ordered and the approximate arriving date and the address. So they have the price. They have a lot of details that they should be actually knowing and that a spam email would not know.} Therefore, based on our data, the following relationships exist: \\
\begin{mdframed}

\textbf{H5.1 - H5.2:}  Perceived adequacy of the [email length, granularity of information] positively affects perceived email legitimacy
\end{mdframed}

\subsection{H6: Effects of     ``trust for  email links"   on  ``perceived email legitimacy"} \label{URLS}
%Several previous studies have investigated users' phishing URL susceptibility without the email context \cite{althobaiti2021don,CHI2020}. Furthermore, even the limited number of studies conducted in the email context do not pay much attention to email links. The reason for this lack of attention can be attributed to the nature of user study designs that have been conducted in the phishing email domain.   As explained in Section~\ref{relatedwork}, most  studies conducted in this domain are based on images of emails or nudge users to mouse over the link during experiment; hence limiting the capacity to provide deeper insights into  how people determine the trustworthiness of email links.

Our results reveal that the absence of email links tends to create a false sense of security in email recipients [\textcolor{blue}{see H6.1}] \blockquote[P18--P]{They don't want me to click on something right. I will just use regular reply to reply to this email}. Most of the time, our participants did not realize that emails without links could also be unsafe as there is a possibility of exposing their personal information by replying to those emails or spreading malware by downloading attachments in phishing emails. 
On the other hand, when there are links in emails, the trust for those links plays a pivotal role in people's intention to respond. Some participants explained that they would trust links that appear to be non-mandatory [\textcolor{blue}{see H6.2}]~\blockquote[P19--P]{It is not asking to do anything particularly. This creates a  huge trust in me. It is not asking me to click a link}. %This is mainly because the participants feel that a phishing attacker would always want to make the email recipient feel that it is mandatory to click on the provided email links in order to achieve their goals.  

We also observed that participants come to conclusions about the trustworthiness of the email links based on the link text and appearance [\textcolor{blue}{see H6.3}]. Some participants explained that they prefer to see the full URL specified in the email body rather than having an alternative text or an image, not realizing that the actual destination can be different from the URL text specified in the email body \blockquote[P04--P]{They are not showing the URL just have the word here [link text]. Why do they have to hide the URL. It's probably because it's a fake website}. Some looked at the button's appearance and made conclusions about the link's legitimacy.
Some even believed that if a link text starts with HTTPS, it simply means a secure connection and the link is trustworthy \blockquote[P18--L]{They [links] usually start with HTTPS. If there's no HTTPS, I might be a little concerned}.  

We only observed a few participants who determined the link destination correctly by hovering over or copying the link address. However, even they made incorrect judgments about link destinations as a result of a lack of understanding of URL structures. For example, some were satisfied seeing the organization's name anywhere in the destination URL. These findings are in line with previous work \cite{CHI2020} on phishing URLs that explained that people are strongly biased towards answering that a URL would lead to the website of the organization whose name appeared in the URL, regardless of its position in the URL structure   [\textcolor{blue}{see H6.4}] \blockquote[P18--P]{I am just looking at the bottom left in the corner. As it has `confirmation', `account', and `security', yeah, it does sound familiar}. Some even make decisions on the link legitimacy based on the similarity of the  URL mentioned in the link text and the destination URL [\textcolor{blue}{see H6.5}]. Some had doubts about the email links when they observed unfamiliar text or numbers in the destination URL [\textcolor{blue}{see H6.6}] \blockquote[P18--P]{It [destination URL] looks really weird to me. Those numbers, I don't know what exactly that means. It makes me feel very uncomfortable clicking on it. Probably, it could contain a lot of information in those numbers. But I don't know what it is}.
%I just look at the bottom left corner there, the link there. It doesn't really hook out with the link here. So, I think its look real at first glance but like when looks at that to the bottom left corner, it does not match-up. I feel it's a fake. 
Therefore, based on our data, the following relationships exist: \\

% \vspace{1em}
\begin{mdframed}

\textbf{H6.1 - H6.5:}  [Absence of email links, Perceived non-mandatory nature of the links,  Perceived trust about links based on the link text and appearance, Appearance of the organization name in the destination URL, Similarity of the URL specified in the link text and the destination URL] positively affects perceived email legitimacy \\
 \textbf{H6.6:}  Appearance of unfamiliar text or numbers  in the destination URL   negatively affects perceived email legitimacy
\end{mdframed}

\subsection{H7: Effects of ``perceived sense of  security from  auxiliary security content"  on ``perceived email legitimacy"} \label{auxiliary}
%Our findings reveal that people tend to develop a sense of security about emails based on the available auxiliary security content. %Unfortunately, to the best of our knowledge, previous research has not shed light on this aspect before.
Our data revealed that people feel a sense of security when an email indicates that it has been scanned by an external scanning tool [\textcolor{blue}{see H7.1}]. One participant explained that he feels he could trust emails that say external tools have scanned the content even when he does not understand  the role of a scanning tool  \blockquote[P18--L]{But yeah, this message here [message saying that the email is scanned for malware by force point] might probably make me loose guard a bit}. On the other hand, we observed that participants tend to trust emails after reading the information provided in the email footer [\textcolor{blue}{see H7.2}]. For example, if the email footer provides reasons why they got the email or organization information such as ABN, or copyright information, then people tend to believe they have received a legitimate email \blockquote[P17--P]{Just having that claim at the bottom [footer] explaining why I got this email. That makes me confident enough to open up that straight away}. Some believe that information contained in the email footer about the company can be easily verified through the internet, not realizing that attackers could also obtain that information without much difficulty. Some even analyzed the phishing education, training, and awareness messages provided in the email [\textcolor{blue}{see H7.3}]. They tend to believe that if an email provides users with anti-phishing education,  there is less possibility for it to be a phishing attack \blockquote[P18--L]{Fake emails usually don't give you methods to identify fake emails}. \\

% \vspace{1em}
\begin{mdframed}

\textbf{H7.1 - H7.3:}  Perceived sense of security based on [information on external scanning tools specified in the email,  email footer information,  in-email security education, and awareness] positively affects perceived email legitimacy
\end{mdframed}

\subsection{H8: Effects of ``previous phishing  experiences"  on ``perceived email legitimacy"} \label{previousphishing}
Our data reveal that past phishing email encounters could make people suspicious about future emails [\textcolor{blue}{see H8.1}] %For example, one participant explained that she would never click on any email link as she has been a victim of phishing emails before 
\blockquote[P1--L]{No, I will never click email link because I have learned a lesson}. People can self-learn correct as well as incorrect strategies for detecting phishing emails based on phishing emails they have seen before [\textcolor{blue}{see H8.2}]. For example,  one participant explained his suspicion over a  user name of the sender's email address as he has seen the same user name in phishing emails before \blockquote[P14--L]{I mean so far with my experience ... I have seen this webmaster, this kind of things. It is generally used by falsified email thing}. 
Several participants indicated that they had received some form of formal education related to phishing, and as a result, they are aware of some strategies to identify phishing emails [\textcolor{blue}{see H8.3}] \blockquote[P01--P]{I also check the punctuation because I've learned from my computer class that phishing emails have misspellings and issues with punctuation}.  \\

% \vspace{1em}
\begin{mdframed}
\textbf{H8.1:} Suspicion for email based on previous phishing email encounters negatively affects perceived email legitimacy \\
\textbf{H8.2 - H8.3}   Trust perceived by applying [self-learned correct and incorrect strategies from past phishing encounters, strategies learned from formal education and training] to detect phishing emails  positively affects perceived email legitimacy
\end{mdframed}

% \label{legitimacy}

\subsection{H9: Effects of ``perceived email legitimacy" on ``intention to respond  an email" } \label{legitimacy}
Our data reveal that often participants decide to click on email links, reply to emails, or download attachments when they perceive an email as legitimate [\textcolor{blue}{see H9}]. For example, a participant, after determining a given phishing email is legitimate, said he will reply to the email immediately as he trusts that email  \blockquote[P13--P]{Yeah, this one's legitimate ... So I will reply like like Hi I'm Martin. so I'll be fine with this.}. Therefore, based on our data, the
following relationship exist
\\

\begin{mdframed}
\textbf{H9:} Perceived email legitimacy positively affects the intention to respond % by clicking/replying/downloading attachments 

\end{mdframed}

\subsection{H10: Effects of ``perceived email legitimacy" on ``intention to validate the email"} \label{legitimacy_validate}
%A large number of previous studies focus on people's email legitimacy judgements~\cite{sarno2021so, jones2019email, wang2016overconfidence, wen2019hack, reinheimer2020investigation}. The main assumption in these studies is that phishing susceptibility can be avoided if users can correctly identify the legitimacy of emails.  
%Confirming this, o
Our data provides evidence that when people have doubts about the email's legitimacy, they may want to validate the email further before deciding on the final response [\textcolor{blue}{see H10}].   In this instance, their email response decision depends on how much trust they can develop about the email legitimacy based on the validation technique instead of the originally perceived email legitimacy  \blockquote[P01--P]{I get a feeling that this might not be a true email. So, first, the action I will be taking is to go to the website and make a call to them}. Details on the types of validation techniques participants described and their effects on the intentions to respond to an email are described in Section~\ref{validationsec}. Therefore, based on our data, the following relationships exist: \\

% \vspace{1em}
\begin{mdframed}
\textbf{H10:} Perceived email legitimacy negatively  affects the intention to validate the email

\end{mdframed}

\subsection{H11, H12: Effects of ``perceived email legitimacy" on ``emotional attachments" } \label{legitimacy_emotional}
We observed situations where participants get furious or frustrated after perceiving an email as phishing. These specific emotions will then drive their response behaviors later.  For example, sometimes, participants expressed their desire to avoid the current email  [\textcolor{blue}{see H11}] as well as future emails [\textcolor{blue}{see H12}] based on anger/frustration resulting from the identification of a phishing email. We provide detailed explanations on how these emotions drive their email response behaviors in Section~\ref{emotionsection}. Therefore, based on our data, the
following relationships exist\\

\begin{mdframed}
\textbf{H11 -  H12:} Perceived email legitimacy negatively  affects [motivation to avoid the current email based on anger/frustration resulting from the identification of a phishing email,  motivation to avoid future phishing emails based on anger/frustration resulting from the identification of a phishing email]

\end{mdframed}

\subsection{H13: Effects of ``intention to validate the email" on ``intention to respond an email" } \label{validationsec}

As explained in Section~\ref{legitimacy_validate},  our findings reveal that email validation is triggered by doubts about email legitimacy. The trust perceived after validation positively influences the intention to respond. %Unfortunately, none of the previous studies on phishing susceptibility shed light on email validation as a part of people's email response decision-making process. 
Unfortunately,  our data revealed that while some of the email validation techniques people use are  safe, others are unsafe. 

Some participants search the internet to look for information and logos specified in the email [\textcolor{blue}{see H13.1}]. They seem to be satisfied if they can find that information online, not realizing the attackers could obtain the same information online when crafting phishing emails \blockquote[P18--L]{I can’t remember the color of the bank logo. I am going to search it up [on the internet]. I am guessing this is the exact same one here [on the internet] as the one here [in the email]}. On the other hand, some participants explained that they would separately log into their accounts to check the information given in the email, which is a much safer option [\textcolor{blue}{see H13.2}] \blockquote[P15--P]{It seems like a bit of a risk to click it, if you want to check Facebook I would instead just go directly to the Facebook website}. 
In the meantime, some participants wanted to check the information by contacting the relevant person/organization. Some choose to call the phone numbers given the email, not realizing that they could be calling phone scammers [\textcolor{blue}{see H13.3}], and others choose to call the phone numbers found on the internet [\textcolor{blue}{see H13.4}]. Others would physically visit the entity (e.g., bank) to verify the email [\textcolor{blue}{see H13.5}]. A participant even explained that she would ask a friend for a second opinion on the legitimacy of the email  [\textcolor{blue}{see H13.6}] \blockquote[P15--P]{It ticks all of my boxes ... But I still get the feeling that maybe the timing is a little bit off. What I will do is ask a friend}.  
We also encountered situations where participants wanted to verify the information by replying to emails, not realizing that it is unsafe to reply to a phishing email [\textcolor{blue}{see H13.7}]. On the other hand, some participants explained that their strategy to verify doubtful emails is to click on the given links and carefully observe the redirecting process or the landing page [\textcolor{blue}{see H13.8}] \blockquote[P18--P]{Whenever there's a weird thing, there's a phishing link, you click on it, and they show like redirecting. Then another website will pop up. So I'll just close the tab at that point.} In summary, the following relationships can exist. \\

% % \vspace{1em}
\begin{mdframed}
 \textbf{H13.1 - H13.6, H13.8:}  Perceived trust  [by validating email content against the  information on the internet, by validating the email content against the information on personal accounts, by calling phone numbers obtained  from the email by calling phone numbers sourced from the internet, by physically visiting the entity, by consulting others, based on the landing pages/process]  positively affects the intention to respond %by  clicking/replying/downloading attachments.
 \\
  \textbf{H13.7:}  Feeling safe to validate  the email  by replying  to the email positively affects the intention to respond %by  clicking/replying/downloading attachments.
\end{mdframed}

% % \vspace{1em}

\subsection{H14: Effects of ``personal habits" on ``intention to respond to an email"} \label{habits}
%A habit is a form of routinized, automatic behavior that occurs without making a conscious decision about the action. 
%Previous work has argued that habits could increase phishing susceptibility because employees are dealing with emails daily, and most emails received in organizations are not harmful; responding to emails has become a habitual behavior \cite{shahbaznezhad2021employees, vishwanath2011people}. However, 

Our findings reveal that  user habits  can affect their intention to respond negatively and positively. %positively as well as negatively affect the user's intention to respond to emails by clicking/replying/downloading attachments. 
We observed several occasions when people make hasty decisions  regarding email responses [\textcolor{blue}{see H14.1}]. There are several reasons; some generally do not read the emails they get and believe it is easy to click on links without much thinking \blockquote[P06--P]{It's in general when someone shares a document with me? I  click it and look}. Some get stressed whenever there is an unattended email notification in their inbox. On the other hand, some participants indicated they would think it is much easier to click on the link during a busy day \blockquote[P15--P]{If I was really busy, I would just click on the link}. 
% Another important fact that affects email response behaviors is the relative importance given to the received email by the user [\textcolor{blue}{see H13.2}]. 
Some  explained that they would consider their priorities and the relative importance of the email if they are having a busy day [\textcolor{blue}{see H14.2}]: \blockquote[P08-L] {It will depend upon  my priorities. If I'm free, I'll open it. If not, I will suspend it. Or I may ignore it completely}.

On the other hand, some seem to completely trust in anti-virus software installed on the computer and tend to transfer the responsibility of detecting and filtering phishing emails to them [\textcolor{blue}{see H14.3}] \blockquote[P14--L]{There is something [anti virus software] to take care of this}. We found another set of users who avoid clicking on links in the email client and use the relevant apps installed in their mobiles to take any required actions [\textcolor{blue}{see H14.4}]. They use those mobile apps as they consider it to be the most convenient option for them: \blockquote[P14--L] {It's easier. The main thing I use Facebook for is Messenger. So it's just there}. Even though these users do not use mobile apps, considering safety reasons, their habit of using the mobile app for convenience reduces their possibility of falling prey to phishing emails.

Some participants explained that  they are generally suspicious about any email that they receive in their inbox  and  avoid responding to any email even when they feel the email is legitimate [\textcolor{blue}{see H14.5}]. Furthermore, some are extra cautious about specific types of emails (e.g., alert/banking emails) [\textcolor{blue}{see H14.6}]. They feel that attackers launch their attacks usually through those types of emails; hence do not want to respond to those types of emails through the email app. Instead, they may use the mobile app or separately log in to the website to carry out any required actions. \\
%\blockquote[P16--P]{I don't know if I will download the PDF. I just don't usually do that. I didn't feel safe then downloading PDFs through email}. Therefore, based on our data, the following relationships exist:

% \vspace{1em}
\begin{mdframed}
 \textbf{H14.1 - H14.3:} [Hasty decision-making, Relative importance given to the email,   Complete trust on anti-virus software]  positively affects the intention to respond %by clicking/replying/downloading attachments 
 \\
  \textbf{H14.4 - H14.6:} [The use of the mobile app for convenience, Skepticality  about any email,  Extra vigilance about alert/banking emails]  negatively affects the intention to respond %by  clicking/replying/downloading attachments  
\end{mdframed}

\subsection{H15: Effects of ``emotional attachments" on ``intention to respond an email" } \label{emotionsection}

%Previous research has highlighted the importance of managing emotions in phishing prevention~\cite{abroshan2021covid, ferreira2019persuasion, lefranc2019factors, workman2008wisecrackers}. However, there are still limited insights into how different types of emotions influence people's email response behaviors. Our findings help to reduce this research gap. 

Our data reveal that people could respond to emails based on emotions even without considering email legitimacy or overlooking judgments they made about email legitimacy. %We explain how diverse emotions drive people's email response behaviors below.    
More specifically, people could be driven to respond to emails when they are happy or excited to receive those emails  [\textcolor{blue}{see H15.1}]. For example, we observed that when participants are offered a job, they are  keen to instantly click on the  links to  view more details  or accept the offer \blockquote[P01--P]{I'm  so happy about the offer. I   will respond soon, I'll accept the offer}. Some emails make people curious, especially if some information is directly not visible in the email  [\textcolor{blue}{see H15.2}]. Curiosity causes people to respond to emails faster  \blockquote[P01--P]{If I am curious to know  more details, I would  go and click order details and see all the details}. Similarly, fear of losing assets, information, and access to accounts could drive people to respond to emails immediately without a conscious decision of the action [\textcolor{blue}{see H15.3}] \blockquote[P03--P]{How has somebody has added this email to my account. This worries me because if my account is linked to PayPal,  all my money will be easily transferred. I would click on this link}.

In our data, anxiety related to work-related priorities and relationships is a very clear driver that made participants  want to respond to emails even  where they doubted the legitimacy [\textcolor{blue}{see H15.4}]. For example, when the participants saw an email from their supervisor, they  wanted to respond immediately \blockquote[P16--P]{I wouldn't question ... The supervisor has, you know, superiority. So, it feels like I have to get this resolved today as I  want to impress}. On the other hand, we observed mixed reactions when participants believed that they received an email from their family. For example, at times, they explained the desire to respond to such emails immediately as they feel it is important to attend to family matters urgently [\textcolor{blue}{see H15.5}]. On the other hand, some may take a more relaxed approach to respond personal emails as they feel that they can always attend to such emails later without any repercussions  [\textcolor{blue}{see H15.6}] \blockquote[P05--P]{It's quite informal, and I feel like I could deal with it at a later without any repercussion}. %so I could even meet my wife a little bit later

In Section~\ref{legitimacy}, we described how people could get angry or frustrated after discovering a phishing email in their inbox. We observed that anger or frustration could lead to mixed reactions in terms of email response. In such situations, mostly people tend to delete or ignore the email  [\textcolor{blue}{see H15.7}]. \blockquote[P19--P]{I feel this is more like spam. I think I probably just deleted it}. On the other hand, we observed situations where participants get frustrated after concluding that they have received a phishing email, clicking on the email links with the intention of avoiding future phishing encounters [\textcolor{blue}{see H15.8}]. For example, a participant explained that he wanted to click on the unsubscribe link to make sure he would not receive such phishing emails from the same sender again \blockquote[P06--P]{l want to click here ... I know that this is a scam email ... But I want to unsubscribe from this newsletter}. 
Therefore, based on our data, the following relationships exist: \\

%14.8 missing
% \vspace{1em}
\begin{mdframed}
\textbf{H15.1 -15.5, 15.8:} [Happiness and excitement, Curiosity, Fear of losing assets, information and access to accounts, Anxiety  about maintaining work priorities and relationships, Anxiety to maintain personal relationships, Motivation to avoid future phishing emails based on anger or frustration] positively affects the intention to respond %by clicking/replying/downloading attachments 
\\
\textbf{H15.6 -15.7:} [Relaxed approach to maintaining personal relationships,  Motivation to avoid current email based on anger or frustration] negatively affects the intention to respond %by  clicking/replying/downloading attachments 
\end{mdframed}

\section{Limitations}
The theoretical model presented in this paper emerged from the qualitative data we collected through in-depth think-aloud sessions and interviews.  We reached theoretical saturation (Figure~\ref{saturation}) which indicates adding more participants from the same group will not influence the model. Furthermore, as explained in Section~\ref{dataanalysis}, we ensured the completeness and accuracy of the model in terms of the collected data through negative case analysis. 
However, it is also important to note that qualitative studies are inherently interpretive, and the findings are based on the studied context, thus challenging to generalize.  Furthermore, as explained in Section~\ref{sec:results}, as this study is an exploration into individuals’ decision-making processes 
that influence their email response decisions, all concepts and relationships that emerged from the qualitative data
(involving 19 participants where each participant went through 12 emails during the study) are included in the model regardless of their frequencies.  Future studies could validate the model and also evaluate the significance of the   elements of the user’s email response decision-making process and their relationships uncovered through our data using large-scale surveys considering diverse demographics, specific phishing-attacks, and real-life settings.

Our participants were university students; who are usually more tech-savvy and have a higher critical thinking ability \cite{williams2019persuasive, liu2020effects}. Previous studies also have revealed demographics like education would not influence or correlate to phishing susceptibility because an individual's knowledge does not reflect on behavior~\cite{liu2020effects}. Nevertheless, similar to many other studies on phishing conducted using a sample from a university population, we cannot compare or make conclusions about other populations with different characteristics.

As explained in Section~\ref{methods}, instead of surveys or retrospective interviewing, a role-play-based think-aloud method is better suited to achieve the aim of this research. Although a  role-play scenario lacks real-world validity as users do not actually receive the emails in their usual inbox, the role-play method provided us advantages over conducting the study using real phishing emails. By the analysis of both phishing and legitimate emails, we were able to obtain much deeper insights into the underlying decision-making processes of people rather than only considering actions performed on phishing emails as in studies with ‘real phishing’ emails.   Additionally, this setup also allowed us to employ twelve emails per participant rather than one phishing email which allowed us to present diverse types of emails.  As a result, the model that emerged from our data enables us to identify general email decision-making flaws that attackers could potentially exploit to launch successful phishing attacks.

Similar to other phishing-related studies that used role-play in their study design~\cite{schiller2023towards,ge2021personal, abroshan2021covid}, we designed the role-play scenario and the emails to be realistic and to match the participants as closely as possible. In fact, we used real phishing emails, real legitimate emails, and real phishing URLs for the study and adapted them to suit the given scenario.   Furthermore,  as mentioned in Section~\ref{methods}, through the simulated email client, the dates and times specified in the email content were automatically adjusted based on the time participants opened those emails. 
%However, it is important to note that the study setup may have primed the participants to behave more vigilantly.   
%Furthermore, 
Similar to previous research \cite{nicholson2017can, wang2016overconfidence}, 
we also told participants the nature of the experiment from the start. Although the researcher clearly explained that the goal is to understand how they make email responses and not to measure how well they perform, the instructions may have primed them to examine emails more vigilantly than they usually do as they could be conscious about their thought processes being monitored by the researchers.  
In addition to fulfilling the ethical and responsible research of informed consent, there can be advantages to participants knowing they are taking part in a phishing-related experiment focusing on their decision-making processes compared to a deceptive study.  
For instance, we observed participants speaking to us effortlessly regarding their decision-making process including but not limited to how  how their previous phishing email exposures and phishing education affected their email response decisions. %Importantly, informing the participants that they were participating in a phishing-related experiment avoided any confusion that may have been caused if they had self-realized the experiment's nature during the study.  

\section{Discussion} \label{discussion}
%Although previous research has highlighted an urgent need to understand why people still fall for phishing emails when designing anti-phishing interventions, there is a wide gap in the literature regarding the ways users make email response decisions and associated flaws that cloud be utilized by successful phishing attacks \cite{wash2020experts, franz2021sok, kirlappos2011security, moreno2017fishing}. Through this research, we extend state of art by providing the 

%Need to include: Previous studies have pointed out that past experience with phishing attacks is a  reason for successfully identifying similar phishing attacks in the future~\cite{downs2006decision, chen2020examination}. While confirming this, the current study provides new and detailed insights into how previous phishing experiences affect people's email legitimacy judgments and responses.

%Phishing is a common type of social engineering attack in which attackers masquerade as trustworthy entities and trick people into disclosing sensitive data, downloading malware, and exposing themselves as well as their organizations to cybercrime. Therefore, it is crucial to take into account the limitations in 
A deeper understanding of human decision-making,  misconceptions, and user assumptions is crucial in the design of anti-phishing education, training, and awareness intervention. Through this work, we extend the state-of-the-art by providing a qualitative user study to propose a theoretical model that interprets different elements and relationships of users' email response decision-making process that influence their email responses. The theoretical model allows us to identify several general email decision-making flaws that attackers could potentially exploit to launch successful phishing attacks. 
%Our study confirms and extends what is known and also provides new insights into what can be considered missing pieces of the puzzle to understand why phishing email attacks still work.  
   Similar to previous studies \cite{parsons2013phishing, williams2019persuasive}, our study highlights that people tend to respond to targetted emails with detailed information contextually aligned to their situations \cite{wash2021knowledge, greene2018user, jaeger2021eyes, greitzer2021experimental, goel2017got, williams2018exploring},  looking professional \cite{greene2018user} or looking familiar \cite{wash2021knowledge,lim2021verbal}. The role of personal characteristics and habits in email decision-making is also highlighted on several occasions  \cite{shahbaznezhad2021employees, ayaburi2019understanding, jones2019email, moody2017phish, greitzer2021experimental}.
  
\subsection{The novelty of the findings}
 
Apart from confirming what is already known,  our study also provides novel or more profound insights into elements of the users'  email response decision-making process.   Moreover, the developed model not only focuses on the high-level categories (e.g.,  personal habits)  but also on the lower concepts  (e.g., trust anti-virus software, the use of mobile apps for convenience, extra vigilance about alert/banking emails) and their relationships with categories or latent variables.  Hence, our model can interpret how different elements of people’s email response decision-making processes could positively and negatively influence their email response behavior, which was lacking in previous literature.  For example, although previous research has hinted at the importance of personal habits \cite{shahbaznezhad2021employees, vishwanath2011people, vishwanath2018suspicion, jayatilaka2021falling} and emotions \cite{abroshan2021covid, ferreira2019persuasion, lefranc2019factors, workman2008wisecrackers, jayatilaka2021falling} in phishing prevention, interpretation of how different types of emotions and personal habits influence response decisions is lacking. Our results help to bridge this gap.  More specifically, our results suggest certain habits (e.g.,  having complete trust in anti-virus software) could increase the possibility of responding to emails, and other habits (e.g.,  extra vigilance about alert/banking emails) could reduce this possibility  (see Section~\ref{habits}). In the phishing context, this implies that certain habits reduce the possibility of falling for phishing while others increase this possibility. However, such detailed insights were missing in previous literature. 
%Furthermore, although previous literature has shed light on  emotions such as anger and happiness in susceptibility to phishing attacks [REF], our study provides further insights (see Section~\ref{emotionsection}) into diverse emotions that could be involved in the email response decision-making process and how they  could even override people's email legitimacy judgments and influence their email response decisions.   

Our findings (see Section~\ref{validationsec}) also suggest that people may not make a final decision on email legitimacy while going through an email and may intend to validate the email before deciding how they want to respond. While some validation techniques people utilize are safe, some techniques  (e.g., searching for information online, calling the phone numbers given in the email, and checking with others) could make them susceptible to potential phishing emails. Furthermore, our study reveals that people could learn correct and incorrect strategies to detect phishing emails from their past phishing encounters (see Section~\ref{previousphishing}) and may even decide on the legitimacy of emails based on auxiliary security content available in emails (see Section~\ref{auxiliary}).

In terms of URLs, several previous studies have investigated how people read and interpret phishing URLs without the email context \cite{althobaiti2021don, CHI2020}. Even the studies conducted in the phishing email context \cite{butavicius2022people, wash2020experts},  do not provide evidence-based insights into users' decision-making processes related to link legitimacy determination.  Our findings help to bridge this gap.  For example,  we found that people's decisions on email legitimacy could be based on the perceived non-mandatory nature of the links and even the appearance of buttons. Furthermore, the presence or absence of email links is also a deciding factor of email legitimacy. Our participants assumed that emails without any links could be safe not realizing that attackers could still use such emails to make people download malware or reveal sensitive information.

  In terms of the email source, in line with previous research \cite{luo2013investigating, moody2017phish, pfeffel2019user, vishwanath2011people, hakim2021phishing, greitzer2021experimental}, our findings reveal how people perceive the sender's legitimacy has a  pivotal in trust they place in a received email (see Section~\ref{senderlegitimacy}). For example,  previous research explains that people are more likely to accept a message when the source presents itself as credible.  Our findings also go in line with the findings in \cite{zheng2022presenting} which found that people do not necessarily have a blind spot for email source details but instead do not properly recognize deception tactics commonly employed by phishing attackers.   However, existing research is limited in explaining how people perceive the legitimacy of the sender by interpreting different parts of the sender's address (e.g.,  based on the display name of the sender address,  sub-domain of the sender address, email addresses specified in the email body, etc.) and confusions they have about determining the sender's legitimacy. Our findings help to bridge this research gap.  

  \subsection{Recommendations for anti-phishing education, training, and awareness intervention design}

  In this section, we discuss the implications of our findings to designers and researchers working in the area of anti-phishing education, training, and awareness interventions.

  \subsubsection{Facilitating to eliminate misconceptions and invalid assumptions}
Our study findings point to several misconceptions and invalid assumptions users have with respect to strategies that phishing attackers use and their capacities. For example, we saw that people often assume that phishing emails always contain URLs; hence, the absence of email links tends to create a false sense of security in email recipients (see H6.1). Some assume that phishing attackers could never have access to detailed information about a specific situation and/or information specific to the participants, such as order details of an Amazon order (see H5.2). Some believe that anti-virus software can protect them from phishing attacks (see H14.3), and some assume that phishing attackers usually launch their attacks (see H14.6) by mainly using specific types of emails (e.g., alert/banking emails).
We saw that such misconceptions and assumptions could drive people to make unsafe response decisions even when they know how to detect phishing emails correctly or have some suspicion about the received email due to other phishing cues that they noticed in the email. Therefore,  anti-phishing education and training should not only focus on guiding people on identifying a phishing email correctly but also eliminate the misconceptions and invalid assumptions they have about the strategies and capacities of phishing attackers. 

\subsubsection{Tailored anti-phishing education, training and awareness}
 Our results provide insights into the diversity and complexity of how people make email responses. For example, our results point to situations where some people struggle to identify the legitimacy of emails, some struggle to validate the emails, and some struggle to take safe actions even after making correct legitimacy judgments. Hence, it's important to focus on targeted training and awareness without a one-size-fits-all approach. We see the value of using personas \cite{neate2019co}, in an organization setting, for designing tailored education, training, and awareness interventions. Personas are fictitious representations of user groups, their goals, and preferences for bridging the gap between designers and the end-users they
are designing for~\cite{quintana2017persona}. While personas have been lauded for their benefits, they are rarely used in the context of phishing prevention. The development of personas requires a deeper understanding of the behavioral and demographic characteristics of the users.  %Additionally, we see value in personalizing the nudging mechanisms considering users' email habits (e.g., the average duration they spend looking at emails) and emotional attachments (see Section 6.3 for  details).  
Hence, we anticipate that the knowledge generated from this study, along with other work that looks into the demographic and personality of people who fall for phishing emails \cite{lawson2020email, jones2019email, ge2021personal}  could serve as a starting point for designing effective tailored anti-phishing interventions in the future. 

  \subsubsection{Shifting  focus  from   \textit{accurate email legitimacy judgements} to \textit{secure email responses}}

Our theoretical model suggests that perceived email legitimacy may not be the sole influencer of email response decisions. However, %most previous user studies \cite{jones2019email, wang2016overconfidence, wen2019hack},
frequently anti-phishing education, training, and awareness interventions \cite{wen2019hack, reinheimer2020investigation, weaver2021training}  often focus only on people's email legitimacy judgments in their study designs and/or in their evaluations. Our results indicate other important factors that should be considered independently or in conjunction with people's email legitimacy judgments to ensure safer email response decisions. For example, our results indicate people could fall prey to phishing even after identifying phishing emails correctly (see H15.8 in Section~\ref{emotionsection}). Therefore, in the future, we expect the focus of the design and evaluation of both research and development on anti-phishing education, training, and awareness interventions to shift from  \textit{accurate email legitimacy judgments} to  \textit{secure email responses}.

  %https://link.springer.com/article/10.1007/s11409-019-09197-5

   \subsubsection{Facilitating  safe  email  validation} 
Given that the trust perceived based on email validation techniques could drive people's email response behavior (see Section~\ref{validationsec}), it is important to make sure people validate emails in safer ways. Unfortunately, our study provides evidence that people use unsafe techniques to validate emails when they have doubts about email legitimacy. To the best of our knowledge, this information was missing in previous literature and is an important part of the puzzle that explains why people fall prey to phishing emails. Therefore, we see value in tools being developed to facilitate employees in organizational settings to validate emails, specifically when they are doubtful about email legitimacy. Jenkins et al.~\cite{jenkins2022phished} proposed initial UI designs for a tool to support people who report phishing so that they can confidentially take appropriate action. 
Althobaiti et al.~\cite{althobaiti2021don} designed a usable report based on the information professionals use to support users in deciding if potential phishing URLs are or are not safe to click on. We anticipate such tools to be integrated into email clients and adapted to provide contextualized just-in-time advice to the users on all aspects of an email that they have doubts. For example, our study provides evidence that users struggle in interpreting the reply-to address and sender address. Users can also be unaware of how to validate the organizational logos visible in emails correctly and, as a result, trust outdated logos based on the results of internet searches they perform. Furthermore, anti-phishing education and training material should also be updated to teach users about safe and unsafe email validation techniques specifically.

  \subsubsection{Giving more prominence to diverse personal habits and emotions in tool design}

Our results provide insights into how different emotions (see Section~\ref{emotionsection}) and personal habits  (see Section~\ref{habits})  can positively and negatively influence people's response behaviors. For example, certain  emotions such as happiness and excitement, anxiety about maintaining work priorities and relationships, and anxiety about maintaining personal relationships positively affect the intention to respond to phishing emails. Emotions can also override people's legitimacy judgments and drive them to make undesirable responses to phishing emails (see H15.8).  As a result, attackers could craft emails to manipulate users' emotions in order to increase the possibility of them falling for those emails. However, despite the importance, previous research work has paid less attention to personal habits and emotions in the design of anti-phishing training, education, and awareness tools~\cite{alshaikh2021applying, goel2017got}. We see value in providing people opportunities to self-assess their habits and emotions and the impact of those on their email response decisions. The knowledge generated by such self-assessment tools can be used to create self-awareness, facilitating people to be more vigilant about their response behaviors and also providing targeted anti-phishing training. Furthermore, the knowledge arising from this research, together with advancements in automatic sentiment and emotion recognition in the email context~\cite{halim2020machine}  could be used to guide users to make safer email responses in the wild.

\subsubsection{Facilitating to assess the validity of self-learned strategies}
Our findings in Section~\ref{validationsec} reveal several inaccurate strategies that people learned through past phishing encounters that they use to detect phishing emails. They could apply those strategies to new emails leading to unsafe email responses to phishing emails (see H8.2). This highlights the need for providing support for users to self-validate strategies they have learned to prevent them from applying those strategies in the future. We see value in providing tool support to users to validate self-learned phishing detection strategies. For example, we anticipate advances in natural language processing, such as semantic similarity \cite{chandrasekaran2021evolution}, can be utilized to develop intelligent systems to process user-defined phishing email detection strategies and provide advice or feedback on whether those strategies are safe.  Furthermore, role-playing simulation games as in \cite{wen2019hack} could be  expanded to focus on correcting the self-learned incorrect strategies revealed through this study.

\section{Conclusion}
In this paper, we investigate in-depth how people make email response decisions while reading their emails. Analysis of the collected qualitative data enabled us to develop a theoretical model that describes how people can be driven to respond to emails by clicking on email links and replying to or downloading attachments based on people's email response decision-making elements and their relationships. Based on an improved understanding of how people make email responses, this study enables us to identify how people can be susceptible to manipulation, even in our controlled experiment environment. We proposed five concrete enhancements to state-of-the-art anti-phishing education, training, and awareness tools to support users in making safe email responses. Among others, we suggest that the goal of anti-phishing education, training, and awareness tools should shift from accurate email legitimacy judgments to secure email responses. Therefore,  we believe our work lays the foundation for improving future anti-phishing interventions to make a significant difference in how we prevent phishing email attacks in the future. 

\section*{Acknowledgment}

This work was carried out while the first author was working at the Centre for Research on Engineering Software Technologies (CREST), University of Adelaide.



%\bibliographystyle{IEEEtranS}

\bibliographystyle{IEEEtranS}
\bibliography{IEEEabrv,sample-base}

\end{document}